\documentclass[preprint]{revtex4}

\usepackage[english]{babel}

\usepackage[letterpaper,top=2cm,bottom=2cm,left=3cm,right=3cm,marginparwidth=1.75cm]{geometry}

\usepackage{amsmath}
\usepackage{physics}
\usepackage{graphicx}
\usepackage{comment}
\usepackage[colorlinks=true, allcolors=blue]{hyperref}

\begin{document}
\title{Dynamic scaling of order parameter fluctuations in model B}

\author{Chandrodoy Chattopadhyay$^1$, Josh Ott$^1$, Thomas Sch\"afer$^1$,  and Vladimir Skokov$^{1,2}$}
\affiliation{$^1$ Department of Physics, North Carolina State University,
Raleigh, NC 27695}
\affiliation{$^2$ RIKEN BNL Research Center, Brookhaven National Laboratory, 
Upton, NY 11973, USA}

\begin{abstract}
We describe numerical simulations of the stochastic diffusion equation
with a conserved charge. We focus on the dynamics in the vicinity of 
a critical point in the Ising universality class. The model we
consider is expected to describe the critical dynamics near a 
possible QCD critical point if the coupling of the order parameter 
to the momentum density of the fluid can be neglected. The simulations 
are performed on a spatial lattice, and the time evolution is 
performed using a Metropolis algorithm. We determine the dynamical 
critical exponent $z\simeq 3.972(2)$, which agrees with predictions 
of the epsilon expansion. We also study non-equilibrium sweeps 
of the reduced temperature and observe approximate Kibble-Zurek
scaling. 

\end{abstract}
\maketitle

\section{Introduction}
\label{sec:intro}

 Understanding the dynamical evolution of fluctuations in the vicinity of 
a critical point in the phase diagram is crucial for interpreting the data 
from the Beam energy Scan (BES) program at the Relativistic Heavy Ion 
Collider (RHIC) at Brookhaven National Laboratory \cite{Stephanov:1998dy,
Bzdak:2019pkr,Bluhm:2020mpc,An:2021wof}. Given the success of the fluid 
dynamic description of heavy ion collisions at RHIC, it is natural to 
assume that the dynamics of fluctuations can also be understood in a fluid 
dynamic framework. The relevant fluid dynamic theories are relativistic 
generalizations of the hydrodynamic models classified by Hohenberg and 
Halperin \cite{Hohenberg:1977ym}. This includes purely relaxational dynamics
(model A) \cite{Schweitzer:2020noq,Schaefer:2022bfm}, the dynamics of a 
conserved order parameter (model B) \cite{Berdnikov:1999ph,Nahrgang:2018afz,
Schweitzer:2021iqk}, a conserved order parameter coupled to the momentum 
density of the fluid (model H) \cite{Son:2004iv}, and the dynamics of 
a chiral order parameter (model G) \cite{Rajagopal:1992qz,Nakano:2011re,
Florio:2021jlx}. Ultimately, the dynamics near a possible critical endpoint 
in the QCD phase diagram is expected to be governed by model H 
\cite{Son:2004iv}, but in the present work we neglect the coupling to 
the momentum density of the fluid and focus on model B. This work builds 
on our earlier study of model A \cite{Schaefer:2022bfm}.

 There are two main approaches to fluid dynamic theories with 
fluctuations. The first is based on performing the noise average 
analytically, and deriving a set of deterministic evolution equations 
for $n$-point functions or $n$-th order cumulants of hydrodynamic 
variables \cite{Mukherjee:2015swa,Akamatsu:2016llw,Stephanov:2017ghc,
Martinez:2018wia,Akamatsu:2018vjr,An:2019osr,An:2019csj,An:2020vri}.
This approach is sometimes described as the hydro-kinetic or Hydro+
method. The second option is to simulate the equations of stochastic 
fluid dynamics \cite{Berges:2009jz,Nahrgang:2018afz,Schweitzer:2020noq,
Schweitzer:2021iqk,Pihan:2022xcl,Schaefer:2022bfm}. The advantage of 
the deterministic method is that the issue of regularization and
renormalization of short-distance noise can be handled analytically. 
The main advantage of stochastic simulations is that there is no need 
for additional approximations. Commonly used approximations include
the truncation of hierarchies of correlation functions, as well as
models for equilibrium two-point functions and relaxation rates. 

 In the present work we follow the second approach and study the 
stochastic dynamics of a conserved order parameter in the vicinity 
of a phase transition in the universality class of the Ising model. 
We employ a simple algorithm based on the Metropolis method
\cite{Florio:2021jlx,Schaefer:2022bfm}. Our goal is to demonstrate 
the efficacy of this algorithm in the case of a fluid dynamic theory 
with a conserved charge. We focus, in particular, on the dynamical 
critical exponent, and on the observation of Kibble-Zurek scaling 
for non-equilibrium sweeps of the parameters of the theory
\cite{Kibble:1980mv,Zurek:1985qw,Zurek:1996sj}. The long-term 
objective is to further generalize the methods presented in 
this work to model H, and to couple the dynamics to a
non-trivial background flow.

\section{Models A and B}
\label{sec:modA}

  Consider a scalar field $\phi(t,\vec{x})$ which evolves according to
a stochastic relaxation equation, 
\begin{align}
\label{modAB}
    \partial_t \phi(t, \vec{x}) = 
    - \hat \Gamma_{A,B}\,  \frac {\delta {\cal H}} {\delta \phi(t, \vec{x})} 
    + \zeta (t, \vec{x}).
\end{align}
For purely relaxational dynamics (model A) $\hat \Gamma_A=\Gamma>0$ is a
constant. In the case of a conserved order parameter (model B) the relaxation
rate is proportional to $\nabla^2$, i.~e.~$\hat \Gamma_B = - \Gamma \nabla^2$.
The Hamiltonian ${\cal H}$ is given by 
\begin{align}
\label{H_Ising}
    {\cal H}  = \int d^dx \left[ \frac{1}{2} (\nabla \phi)^2  
    +  \frac{1}{2} m^2 \phi^2(t,\vec{x}) 
    +   \frac{1}{4} \lambda  \phi^4(t,\vec{x})  
    - h(t,\vec{x}) \phi (t,\vec{x})\right] \, ,
\end{align}
where $m$ is the bare inverse correlation length, $\lambda$ is the self-coupling, 
and $h$ is an external field. The noise term $\zeta (t,\vec{x})$ is a random
field constrained by fluctuation-dissipation relations. This determines the 
correlation function of the noise 
\begin{align}
    \langle \zeta (t, \vec{x}) \zeta (t', \vec{x}') \rangle = 
    2 T\, \hat \Gamma_{A,B}\, \delta(\vec{x}-\vec{x}')\delta(t-t')\, .
\end{align}
In order to perform numerical simulations we discretize the Hamiltonian on a 
lattice with lattice spacing $a$. We use periodic boundary conditions on a cubic 
lattice with sides of length $L=Na$, and we adopt units so that $a=1$. The 
discretized Hamiltonian is 
\begin{align}
\label{H_Ising_2}
{\cal H}  = \sum_{\vec{x}} \left[ \frac{1}{2}  
    \sum_{\mu=1}^d   (\phi(\vec{x}+\hat{\mu}) - \phi(\vec{x}) )^2  
    +  \frac{1}{2} m^2 \phi^2(\vec{x}) +   \frac{1}{4} \lambda  \phi^4(\vec{x})  
    - h \phi (\vec{x})\right] \, , 
\end{align}
where we have suppressed the time argument of the field $\phi(\vec{x})$, and 
the sum over $\vec{x}$ is a sum over integer vectors $\vec{n}$ so that $\vec{x}
=a\vec{n}$. We also define a unit vector $\hat \mu$ in the direction $\mu\in 
\{ 1,\ldots, d \}$, where $d$ is the number of spatial dimensions. In the 
present work we will only consider $d=3$.

 In order to study time evolution we have to determine the change in the 
Hamiltonian for an update of the field. In model A the update is purely 
local. Changing the field $\phi_{\rm old} \to \phi_{\rm new}$ at a fixed 
position $\vec{x}$ leads to 
\begin{align}
\label{Delta_H_modA}
    \Delta {\cal H}  (x)&= 
    d (\phi^2_{\rm new}(\vec{x}) - \phi^2_{\rm old}(\vec{x}))
    -  (\phi_{\rm new}(\vec{x}) - \phi_{\rm old}(\vec{x}))
    \sum_{\mu=1}^d \left( \phi(\vec{x}+\hat{\mu}) + \phi(\vec{x}-\hat{\mu})  
    \right)\notag \\   
    &+\frac{1}{2} m^2 ( \phi^2_{\rm new}(\vec{x}) - \phi^2_{\rm old}(\vec{x}))  
    +   \frac{1}{4} \lambda  ( \phi^4_{\rm new}(\vec{x}) 
    - \phi^4_{\rm old}(\vec{x}))  - h (\phi_{\rm new}(\vec{x}) 
    - \phi_{\rm old}(\vec{x})) \,.
\end{align}
In model B the order parameter is conserved, and an update involves two adjacent 
sites at the position $x$ and $x+\hat \nu$. The change in Hamiltonian is 
\begin{align}
\label{Delta_H_modB}
    \Delta {\cal H} (x, x+\hat \nu) &=  
    \Delta {\cal H}(x)  + \Delta {\cal H} (x+\hat \nu)
 \notag \\ & -   ( \phi_{\rm new}(\vec{x})- \phi_{\rm old}(\vec{x}))
 \left( \phi_{\rm new}(\vec{x}+\hat{\nu}) - \phi_{\rm old}(\vec{x}+\hat{\nu} )\right)\, ,
\end{align}
where, for a conserved field, we have $q\equiv \phi_{\rm new}(\vec{x})- \phi_{\rm old}
(\vec{x}) = - [\phi_{\rm new}(\vec{x}+\hat{\nu}) - \phi_{\rm old}(\vec{x}+\hat{\nu}) ]$,
so that the last term in equ.~(\ref{Delta_H_modB}) is equal to $+q^2$.

\subsection{Model A update}

 Following our earlier work \cite{Schaefer:2022bfm} as well as
the recent publication by Florio et al.~\cite{Florio:2021jlx} we 
study the stochastic evolution using the Metropolis method.  
The basic observation is that one can use a single Metropolis 
step to implement both the diffusive and the stochastic terms 
in the equation of motion. There is some rigorous work on this 
method in the mathematics literature \cite{Gao:2021}, but it has
not been used widely in either mathematics or physics. For some
exceptions, see \cite{Rossky:1978,Roberts:1998} as well as
\cite{Moore:1998zk,Hasenbusch:2019gmx}. The time evolution is 
discretized using a fixed time step size $\Delta t$. We perform 
a checkerboard sweep through the spatial lattice. The checkerboard
consists of even lattice sites (the A lattice) and odd lattice 
sites (the B lattice). For every site $\vec{x}\in\text{A}$ we perform 
a trial update
\begin{align}
\label{modA_trial}
    \phi_{\rm new}(\vec{x}) 
 =  \phi_{\rm old}(\vec{x}) + \sqrt{2\Gamma \Delta t}\, \xi\,.  
\end{align}
Here $\xi$ is a random number drawn from a Gaussian distribution 
with zero mean and variance one. The trial update is accepted
with probability $P={\rm min} (1, e^{-\Delta {\cal H}/T})$. In 
this case, we update the field as
\begin{align}
\label{modA_update}
    \phi(t+\Delta t, \vec{x}) =  \phi_{\rm new}(\vec{x})\,   
\end{align}
If the trial update is rejected then the field is not changed.
This update implies that 
\begin{align}
 \large\langle \phi(t+\Delta t,\vec{x}) -\phi(t,\vec{x}) \large\rangle 
    & = - (\Delta t)\, \Gamma \,\frac{\delta {\cal H}}{\delta\phi}
    + O\left( (\Delta t)^2\right) \, , \\
\large\langle \left[\phi(t+\Delta t,\vec{x}) -\phi(t,\vec{x})
    \right]^2 \large\rangle 
    & =  2(\Delta t)\, \Gamma \, T
    + O\left( (\Delta t)^2\right) \, . 
\end{align}
The first moment ensures that the deterministic part of equ.~(\ref{modAB})
is satisfied, and the second moment reproduces the variance of the 
stochastic force in the equation of motion.

The same procedure is repeated for all points in the B lattice. 
For the nearest neighbor interaction in equ.~(\ref{H_Ising_2})
the updates of all $\vec{x}\in\text{A}$ are independent of each other, 
and can be performed in parallel. The same is true for updates in 
the B lattice. An important property of the procedure is that the 
probability of obtaining a new configuration $(\phi^A_{\rm new},
\phi^B_{\rm new})$, where $\phi^{A,B}$ are the fields on the 
A, B sublattices, only depends on the difference in the Hamiltonian
of the initial and final states
\begin{align}
\label{det_bal}
 P\left( (\phi^A,\phi^B)\to  (\phi^A_{\rm new},\phi^B_{\rm new})\right)   
  \sim \exp\left(-[{\cal H }(\phi^A_{\rm new},\phi^B_{\rm new})
      -{\cal H} (\phi^A,\phi^B) ]/T \right), 
\end{align}
and not on the order in which the updates are performed, or on the 
intermediate values of the fields. We also note that the detailed 
balance condition is satisfied irrespective of the size of the 
time step. This means that the Metropolis algorithm defined by 
equ.~(\ref{modA_trial},\ref{modA_update}) samples the equilibrium
distribution $P[\phi]\sim\exp(-{\cal H}[\phi]/T)$ even if $\Delta t$
is not small. In contrast, if $\Delta t$ is not small then the model A
dynamics in equ.~(\ref{modAB}) is only realized approximately. 
However, because the model A/B equation represents the general low 
energy approximation to relaxational dynamics in the absence/presence
of a conservation law, the leading effect of a non-zero $\Delta t$
is to modify the value of the relaxation rate.

\subsection{Model B update}

\begin{figure}
\centering
\includegraphics[width=0.85\linewidth]{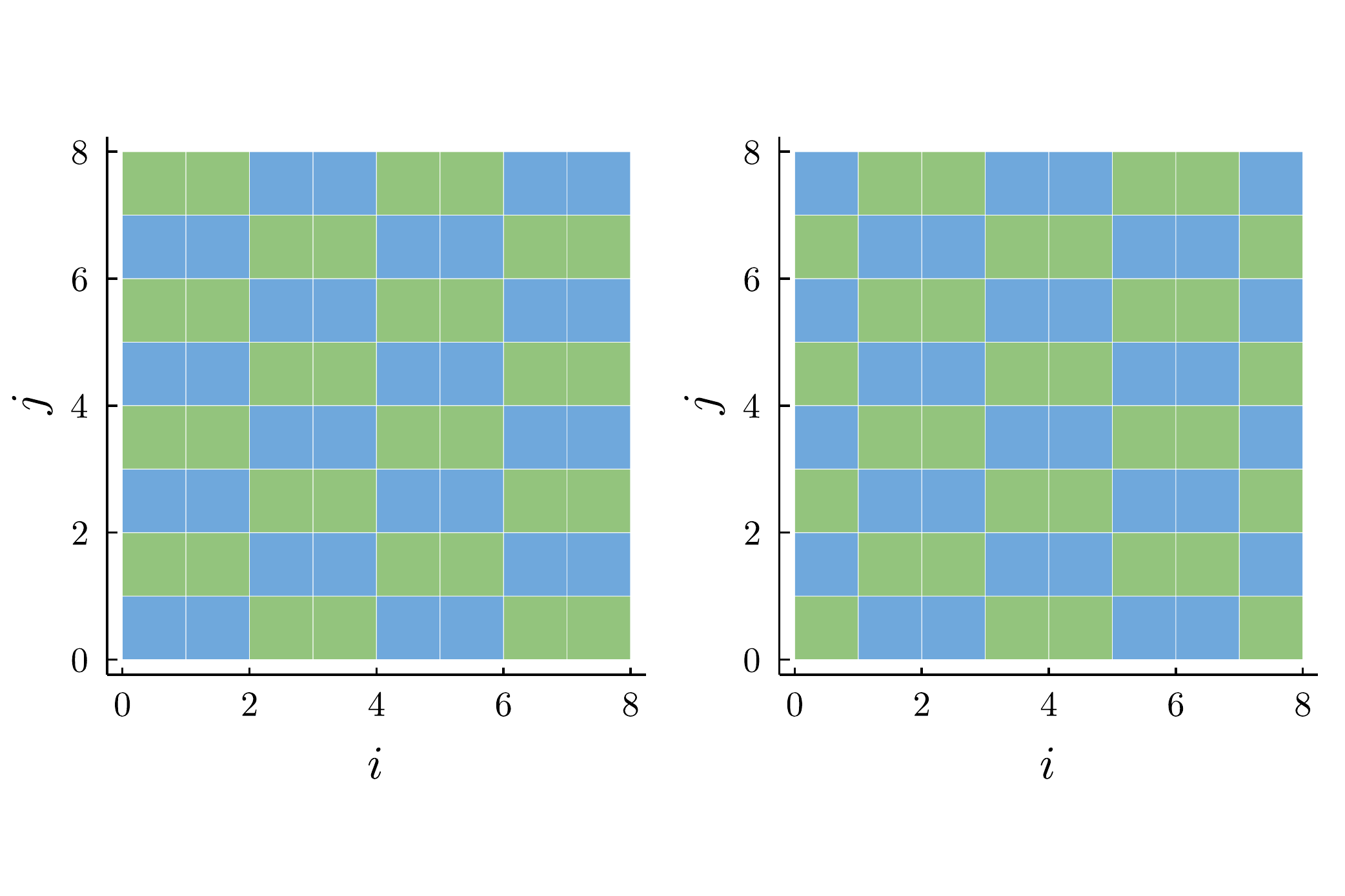}
\caption{Checkerboard algorithm for a conserving (model B) update.
For illustration we show two stages (left and right panel) of the 
algorithm acting on one plane of a three-dimensional lattice of
size $L=8$ labeled by indices $(ijk)$. In the first stage (left panel) 
pairs of green cells are updated by applying a stochastic flux at their 
interface. Blue cells are unmodified. In the $k$-direction (not shown) 
alternate layers have green and blue cells interchanged. In the second 
stage (right panel) the update is applied to a shifted set of cells. 
Stages three and four correspond to exchanging green and blue 
cells relative to stage one and two. Finally, the same pattern
is applied in the $(j,k)$ and $(k,i)$ plane.}
\label{fig:updates}
\end{figure}

In the case of model B we wish to ensure that the update of 
the field respects the conservation law exactly. For this 
purpose we write the equation of motion in the form
\begin{align}
\label{modB}
\partial_t\phi + \vec\nabla\cdot \vec\jmath = 0 \, , 
\hspace{1cm} 
\vec\jmath = -\Gamma \, \vec\nabla\,
   \frac{\delta{\cal H}}{\delta\phi}  + \vec\xi\, , 
\end{align}
with 
\begin{align}
    \langle \xi_i(\vec{x},t)\xi_j(\vec{x}',t')\rangle 
     = \delta_{ij}\Gamma\,  T \delta(\vec{x}-\vec{x'}) 
        \delta(t-t') \, . 
\end{align}
Equation (\ref{modB}) can be integrated over a cell centered
on $\vec{x}=a\vec{n}$. The integral of $\phi$ over a cell 
is $\phi_c(\vec{x})= a^d\phi(\vec{x})$ and in units $a=1$ there
is no need to distinguish between $\phi_c$ and $\phi$. The
evolution equation for $\phi=\phi_c$ is 
\begin{align}
    \partial_t \phi 
     = -\sum_{\mu=1}^d \left( q_\mu^+ - q_\mu^- \right)\, , 
\end{align}
where $q_i^\pm$ are the fluxes through the forward and 
backward faces of the cube centered on $\vec{x}=a\vec{n}$
in the cartesian $i$-direction. We can use the Metropolis algorithm
discussed in the previous section to update the fluxes. A trial
update for the cells $(\vec{x},\vec{x}+\hat{\mu})$ is
\begin{align}
\label{modB_upd}
\begin{array}{rl}
\phi_{\rm new}(\vec{x}) & = \phi_{\rm old}(\vec{x}) + q_\mu\, ,\\
\phi_{\rm new}(\vec{x}+\hat{\mu}) &= \phi_{\rm old}(\vec{x}+\hat{\mu}) 
  - q_\mu\,  ,
\end{array}
\hspace{1cm} q_\mu = \sqrt{2\Gamma\Delta t}\, \xi  \, . 
\end{align}
The update is accepted with probability ${\rm min}(1,e^{-\Delta
{\cal H}})$. 

In the case of the conserving update in equ.~(\ref{modB_upd}) it
is more complicated to construct a checkerboard algorithm. The 
update involves two cells, and for a nearest-neighbor Hamiltonian 
these two cells have $4d-2$ neighbors. We use a fourfold checkerboard
which is repeated in all $d$ directions to update all interfaces
while avoiding interference between neighboring cells. This ensures
that the detailed balance condition equ.~(\ref{det_bal}) is satisfied.

 Two stages of the algorithm are shown in the left and right panel 
of Fig.~\ref{fig:updates}. The figure shows a two-dimensional slice 
in the $(ij)$ plane of a three dimensional lattice labeled by $(xyz)=
(ijk)$. In the two stages shown every pair of green cells is updated 
by applying a stochastic flux at the interface in $i$ direction. The 
figure is replicated in the $k$ direction by alternating green and
blue cells. All these updates can be performed in parallel. Stages
three and four correspond to repeating the procedure with green 
and blue cells interchanged. After these four stages are complete
all interfaces in the $x=i$ direction have been updated. The process is
then repeated for all right-handed permutations of $x$, $y$ and $z$ 
\begin{align}
    (i,j,k) \to (x,y,z),(y,z,x),(z,x,y)\, . 
\end{align}
This completes a full update of the lattice. As in the case of model 
A the update satisfies detailed balance exactly, even for finite $\Delta t$.
However, the diffusion equation is only realized up to correction of 
higher order in $\Delta t$.

\begin{figure}
\centering
\includegraphics[width=0.66\linewidth]{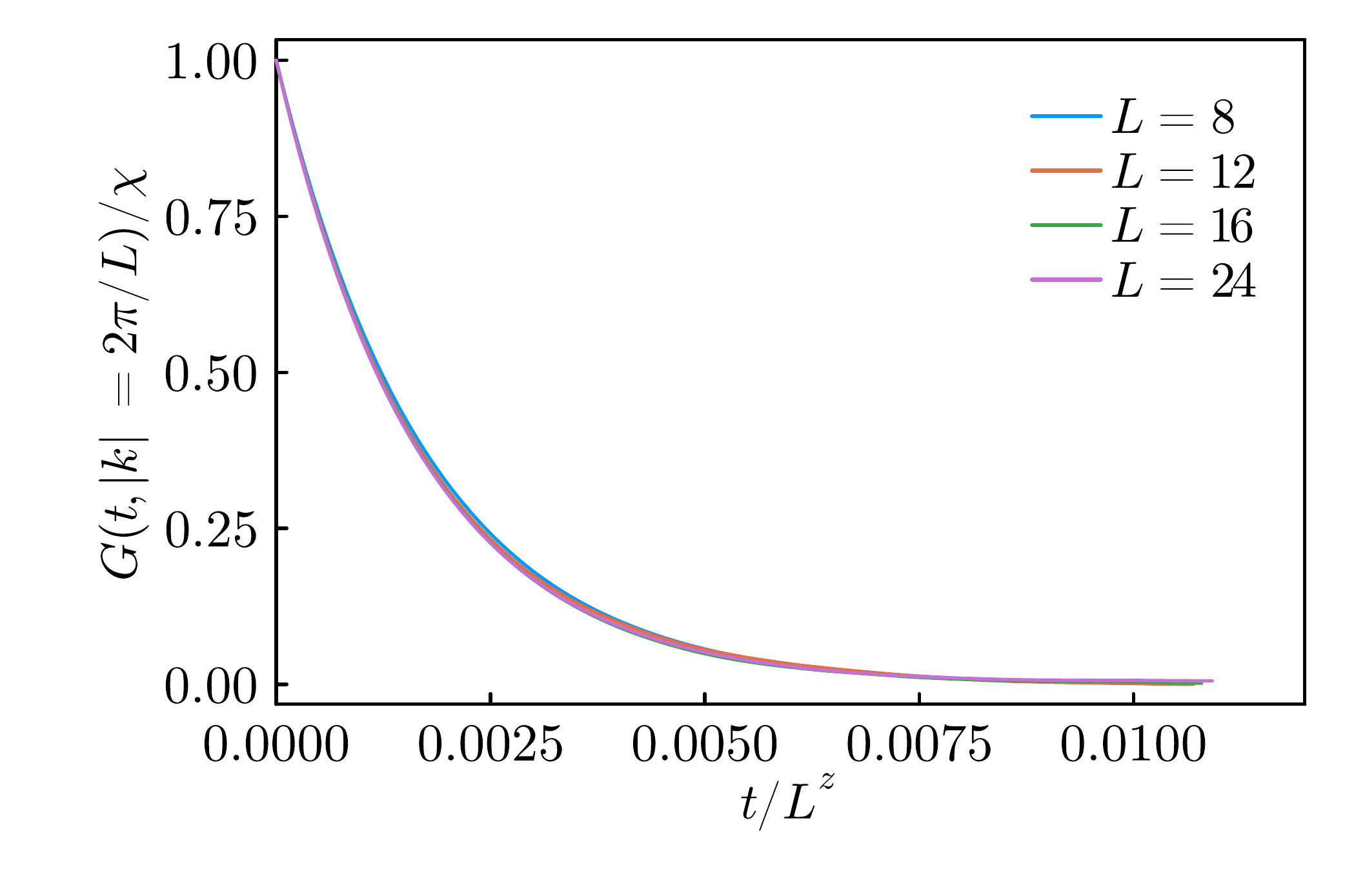}
\caption{Dynamic density-density correlation function $G(t,\vec{k})$ for 
the lowest lattice momentum $\vec{k} = (2\pi/L,0,0)$. We show the results 
obtained in three different volumes with linear size $L=8,12,16,24$. 
Pale bands indicate the statistical error of the data. The time variable 
is scaled by $L^z$ with $z=3.972$. The correlation function is 
normalized to $\chi=G(0,2\pi/L)$.}
\label{fig:cL}
\end{figure}

\section{Numerical Simulations}
\label{sec:statics}

  Numerical simulations of model A dynamics were described in our 
previous work \cite{Schaefer:2022bfm}. In that work we used Binder
cumulants to determine the critical value of $m^2$ at which the second
order phase transition takes place. For $\lambda=4$ we obtained $m_c^2 
= -2.28587(7)$. Since static properties of model B are identical to 
those of model A, this result directly carries over to the present 
work. As we will discuss in more detail below, there is a caveat 
related to the fact that static correlation functions in a finite 
volume are affected by global conservation laws. In particular, in 
our model B simulations the total charge is always zero, whereas the 
total charge fluctuates in simulations of model A. 

 In this section we first study the correlation function directly 
at the critical point $m_c^2$, where the correlation length in the
thermodynamic limit diverges. In a finite volume the correlation 
length is limited by the system size $L$, and we can use this 
dependence to analyze dynamical scaling. We then investigate the 
correlator at $m^2\neq m_c^2$ and extract the correlation length 
and relaxation time away from the critical point. In Sect.~\ref{sec:KZ} 
we study non-equilibrium sweeps, and in this case the correlation 
length is limited by critical slowing down. Note that we have 
set $a=1$, and all lengths are measured in units of the lattice 
spacing. We have also used $\Gamma=1$, and the unit of time is 
given by $a^4/\Gamma$. Finally, we have set $T=1$, and the 
magnetization $\langle\phi\rangle$ is given in units $T/a^{1/2}$.
 
\begin{figure}
\centering
\includegraphics[width=0.66\textwidth]{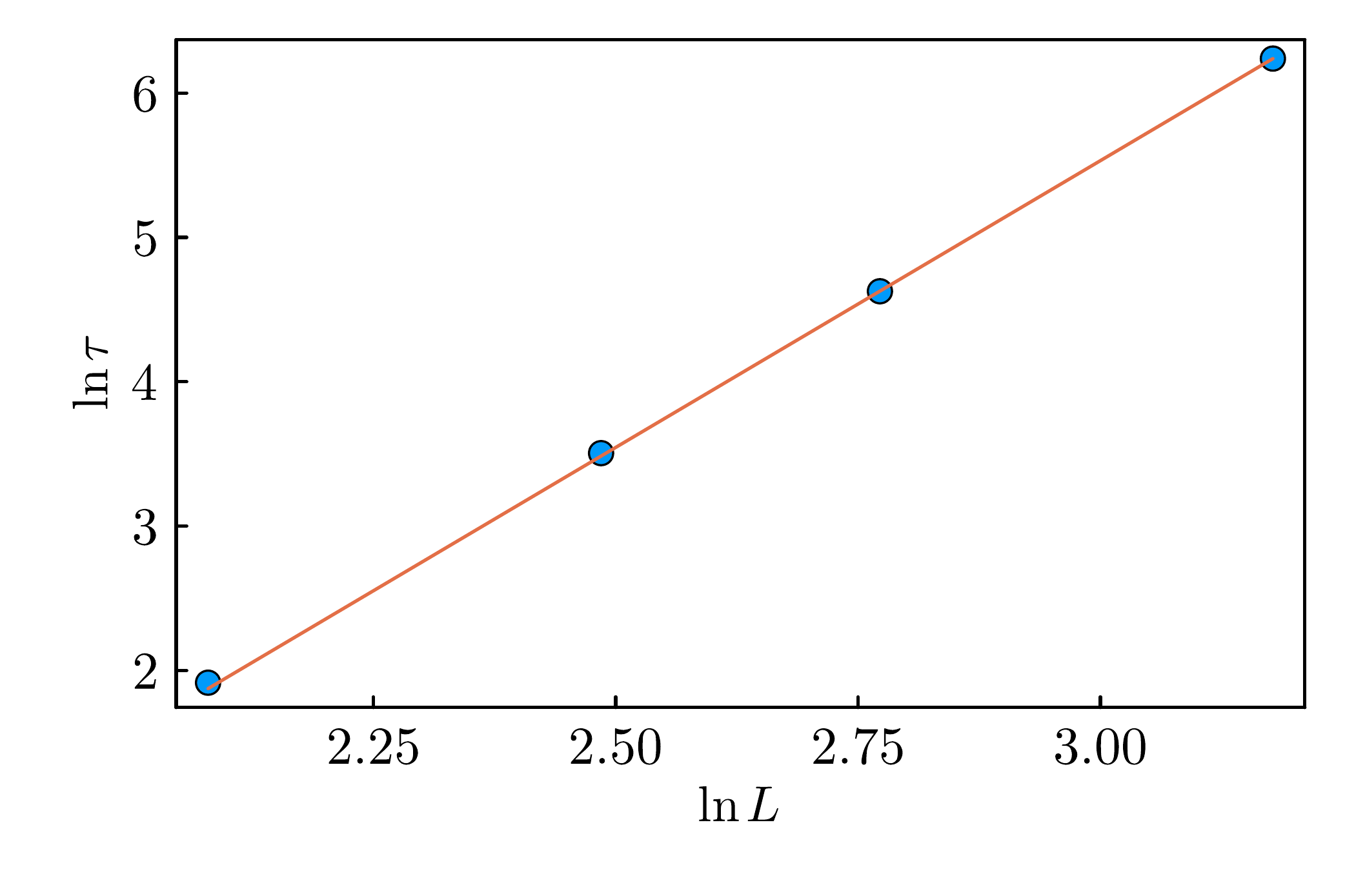}
\caption{Relation between the decay time $\tau$ extracted from the 
correlation function $G(t,|\vec{k}|=2\pi/L)$ and the linear box 
size $L$. The line shows the best fit with $z=3.972$.
%
\label{fig:z_fit}}
\end{figure}

 The dynamic correlation function of the density is defined by 
\begin{align}
G(t,\vec{k}) = \int d^3x\, e^{i\vec{k}\cdot\vec{x}} 
  \langle \phi(0,0)\phi(\vec{x},t) \rangle. 
\end{align}
In Fig.~\ref{fig:cL} we show this function for the lowest non-trivial 
lattice momentum at the critical point for four different lattice volumes. 
Critical scaling predicts that correlation functions obtained in different 
volumes collapse to a universal function if the time argument is scaled 
by $L^z$, where $z$ is the dynamic critical exponent of model B. In 
Fig.~\ref{fig:cL} we show that data collapse occurs for $z=3.972$. A 
more detailed analysis of the dynamic critical exponent is shown in 
Fig.~\ref{fig:z_fit}. For different values of $L$ we show the relaxation 
time extracted from a fit $G(t,\vec{k})\sim e^{-t/\tau}$ where $\vec{k}=
(2\pi/L,0,0)$. We observe that the dependence of $\log(\tau)$ on $\log(L)$ 
is linear, and from the slope of this relation we determine the value 
$z=3.972$. We can get an estimate of finite $L$ corrections by excluding 
the smallest $L$-value, $L=8$, from the fit. In this case we get $z=3.970$,
and we conclude that $z=3.972(2)$. Other sources of error are more difficult 
to quantify. As mentioned above there is a (small) uncertainty in the 
determination of $m_c^2$, and we have not attempted to quantify the 
impact of this error on the measurement of $z$. We can compare our 
result $z=3.972(2)$ to the theoretical prediction of $z=4-\eta$ 
\cite{Hohenberg:1977ym}, where $\eta$ is the static correlation function 
exponent defined by $G(0,|\vec{k}|)\sim 1/|\vec{k}|^{2-\eta}$. In the
$\epsilon$ expansion $\eta=3\epsilon^2/162 \simeq 0.019$ 
\cite{Zinn-Justin:2002ecy}, and in the conformal bootstrap 
$\eta\simeq 0.0363$ \cite{El-Showk:2014dwa,Alday:2015ota}. Based on 
the latter result we expect $z\simeq 3.96$, in good agreement with 
our result. 

 We note that the large value of $z$ implies that configurations in model 
 B are very difficult to thermalize. The effort to update a single 
configuration scales like the volume $V\sim L^3$, and the number of updates
required to thermalize the system scales as $L^z\sim L^4$. This implies
that at the critical point the effort scales roughly as $L^7$. Away from
the critical point stochastic simulations are significantly faster. 

\begin{figure}
\centering
\includegraphics[width=0.66\textwidth]{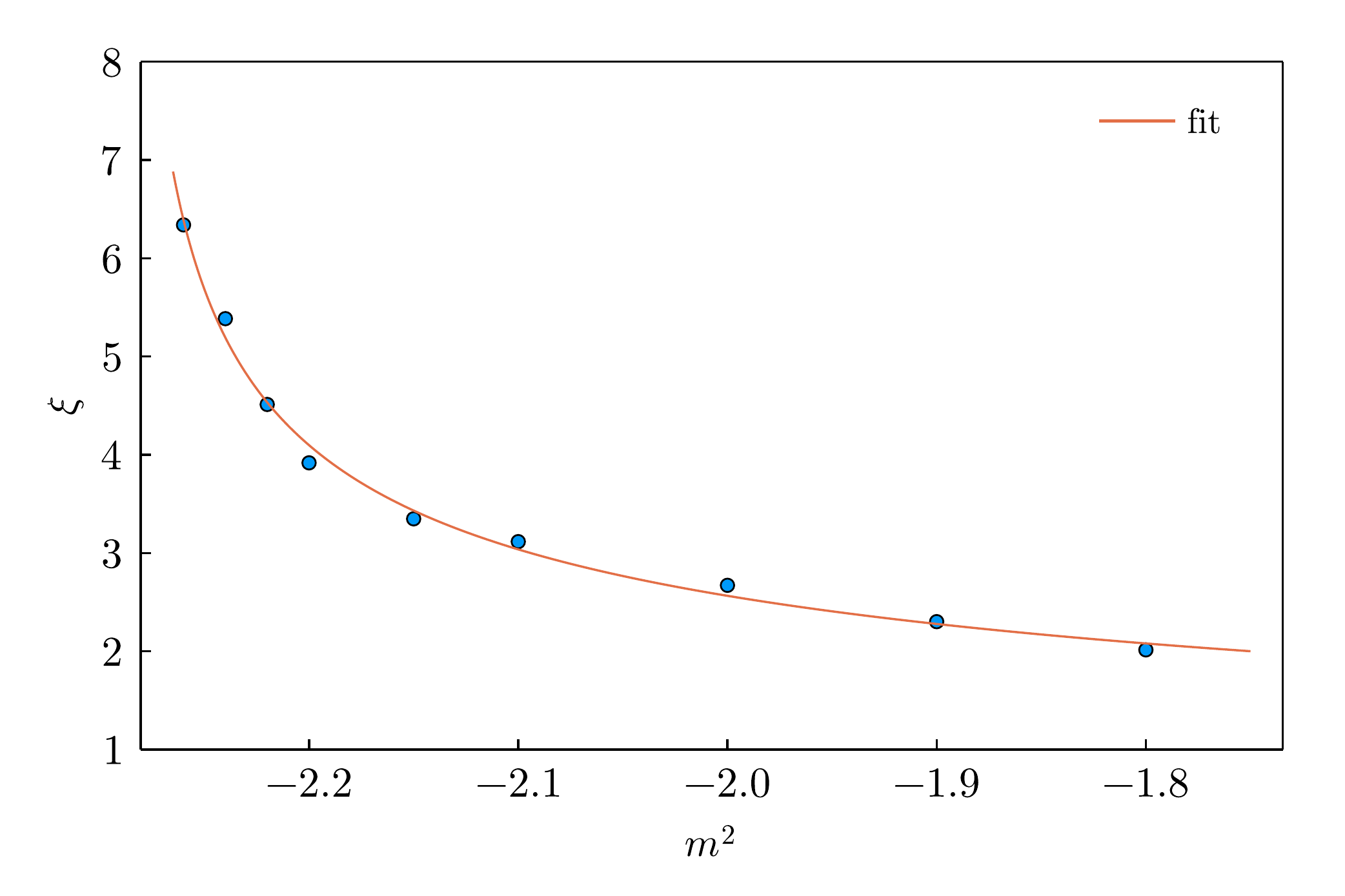}
\caption{Model B correlation length $\xi$ as a function of the bare mass
parameter $m^2$. We also show a fit $\xi\sim 1/(m^2-m_c^2)^\nu$ with
$\nu=0.542$ and $m_c^2=-2.267$.
\label{fig:xi_modB}}
\end{figure}

 We have also studied correlation functions away from the critical point
$m^2=m_c^2$. In Fig.~\ref{fig:xi_modB} we show the correlation length
extracted from a fit to the equal time correlation function $G(0,\vec{x})$.
In particular, we write $G(0,\vec{x}) \sim x^{-1}\exp(-x/\xi)$ with 
$x=|\vec{x}|$ and fit $\xi$ to the measured correlation function in the 
regime $x<L/2$. In practice, the fit window has to be smaller, because 
conservation of total charge implies that the model B correlation function
at $x\sim L/2$ is negative. Indeed, we find that because of finite size 
corrections the model B correlation length in any finite system is smaller
than the one in model A, even though the two theories are governed by the 
same static universality class. We have fitted the correlation lengths
shown in Fig.~\ref{fig:xi_modB} to a power law $\xi\sim 1/(m^2-m_c^2)^\nu$.
We find $m_c^2=-2.267$ and $\nu=0.542$. The value of $m_c^2$ is consistent
with (but much less precise than) the value $m_c^2=-2.28587(7)$ extracted
from the Binder cumulants. Similarly, because of finite size effects, the 
value of $\nu$ is consistent with the expectation $\nu=0.6299(5)$ 
\cite{El-Showk:2014dwa}, but not very precise.

\begin{figure}
\centering
\includegraphics[width=0.65\textwidth]{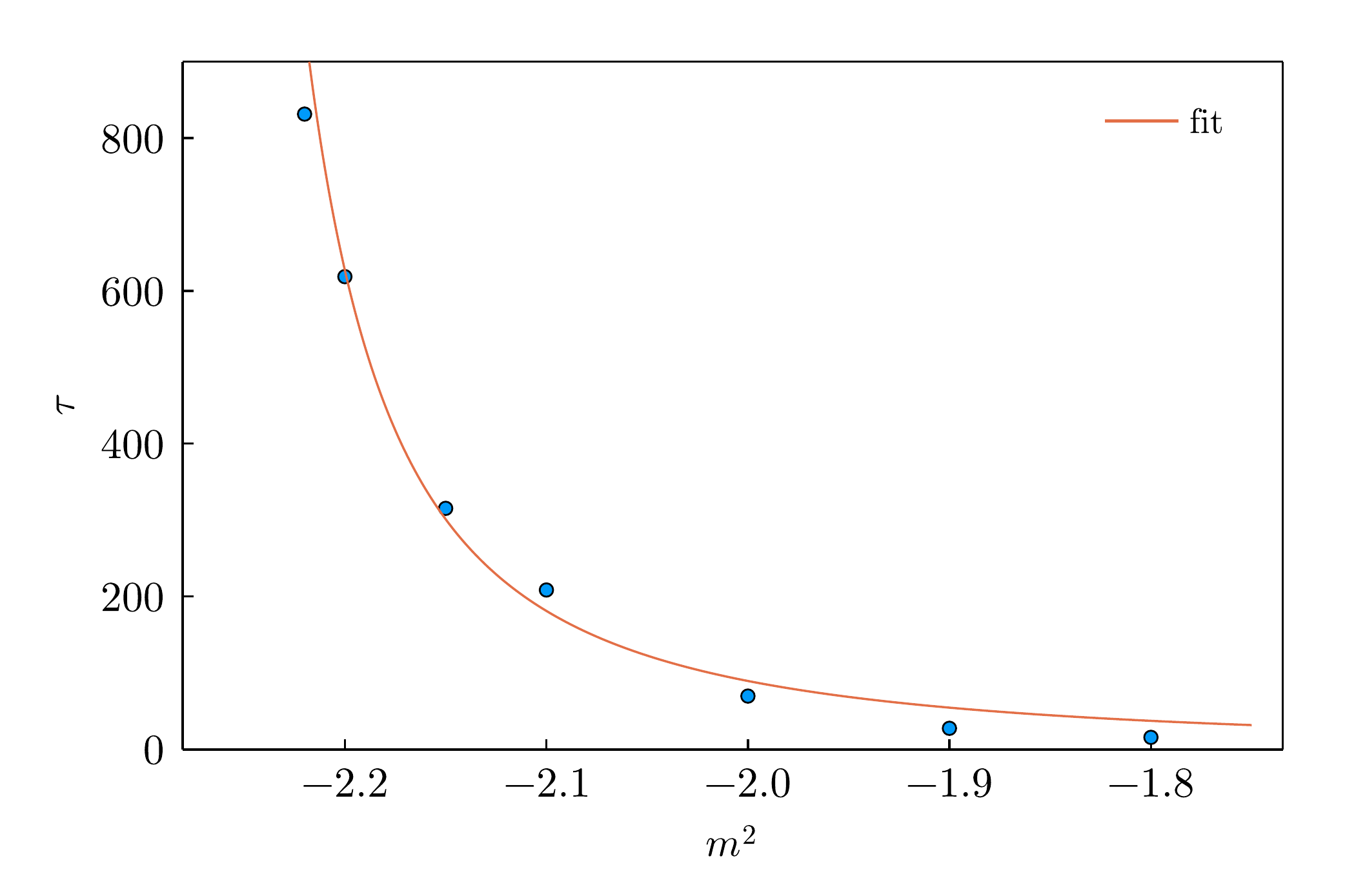}
\caption{Model B relaxation time $\tau$ as a function of the bare mass
parameter. Here, $\tau$ is measured for the lowest non-trivial Fourier
mode $k=2\pi/L$. We also show the best fit $\tau\sim \xi^z$ with 
$z\simeq 3.73$.
\label{fig:tau_modB}}
\end{figure}

 In Fig.~\ref{fig:tau_modB} we show the dependence of the relaxation time
on the bare mass parameter $m^2$. The relaxation time is extracted for
the lowest Fourier mode $k=2\pi/L$ by using a simple exponential fit
$G(t,k=2\pi/L)\sim\exp(-t/\tau)$. Fig.~\ref{fig:tau_modB} clearly 
demonstrates that the variation of the relaxation time with $m^2$
is much more dramatic than that of the correlation length. A simple
fit of the form $\tau\sim\xi^z$ gives $z\simeq 3.73$. This fit is 
consistent with, but not nearly as accurate, as the determination 
of $z$ from finite-size scaling at the critical point $m^2=m_c^2$ 
shown in Fig.~\ref{fig:z_fit}.

\section{Kibble-Zurek scaling}
\label{sec:KZ}

  In practical applications of stochastic diffusion we are often interested
in far-from-equilibrium dynamics. In our previous work on the dynamics of
model A we studied quench dynamics \cite{Schaefer:2022bfm}: We equilibrated
the system in the high temperature phase, and performed an instantaneous
sweep to the critical point. We then studied the evolution of moments of the 
order parameter towards their equilibrium values at the critical point. 

 In this section we study a different situation. We again equilibrate the
system in the high temperature phase, but then sweep towards the critical 
regime at a finite rate. Because of critical slowing down, correlation
functions drop out of equilibrium as the critical point is approached. This 
behavior can be characterized in terms of a time scale, the Kibble-Zurek 
time $\tau_{KZ}$ \cite{Kibble:1980mv,Zurek:1985qw,Zurek:1996sj,Chandran:2012,
Akamatsu:2016llw,Akamatsu:2018vjr}. This is the time at which fluctuations 
on a length scale defined by the instantaneous correlation length $\xi$ 
are no longer in equilibrium. The correlation length at that time is the
Kibble-Zurek length $l_{KZ}$. Consider a sweep that passes the critical 
point at $t=0$. The idea of Kibble-Zurek scaling is that there is a scaling 
window $t,t'\in[-\tau_{KZ},\tau_{KZ}]$ so that the non-equilibrium two-point 
function of an observable ${\cal O}$ satisfies
\begin{align}
G_{\cal O}(t,t',k) = l_{KZ}^{\Delta_{\cal O}}\; 
   g_{\cal O}\left( \frac{t+t'}{2\tau_{KZ}};
            \frac{t-t'}{\tau_{KZ}},kl_{KZ}\right)\, ,
\end{align}
where $g_{\cal O}$ is a universal function and $\Delta_{\cal O}$ is an
anomalous dimension. In the following we will test Kibble-Zurek scaling 
by comparing equal time correlation functions computed for different 
quench rates.

\begin{figure}
\hspace*{0.10\linewidth}
\includegraphics[width=0.66\linewidth]{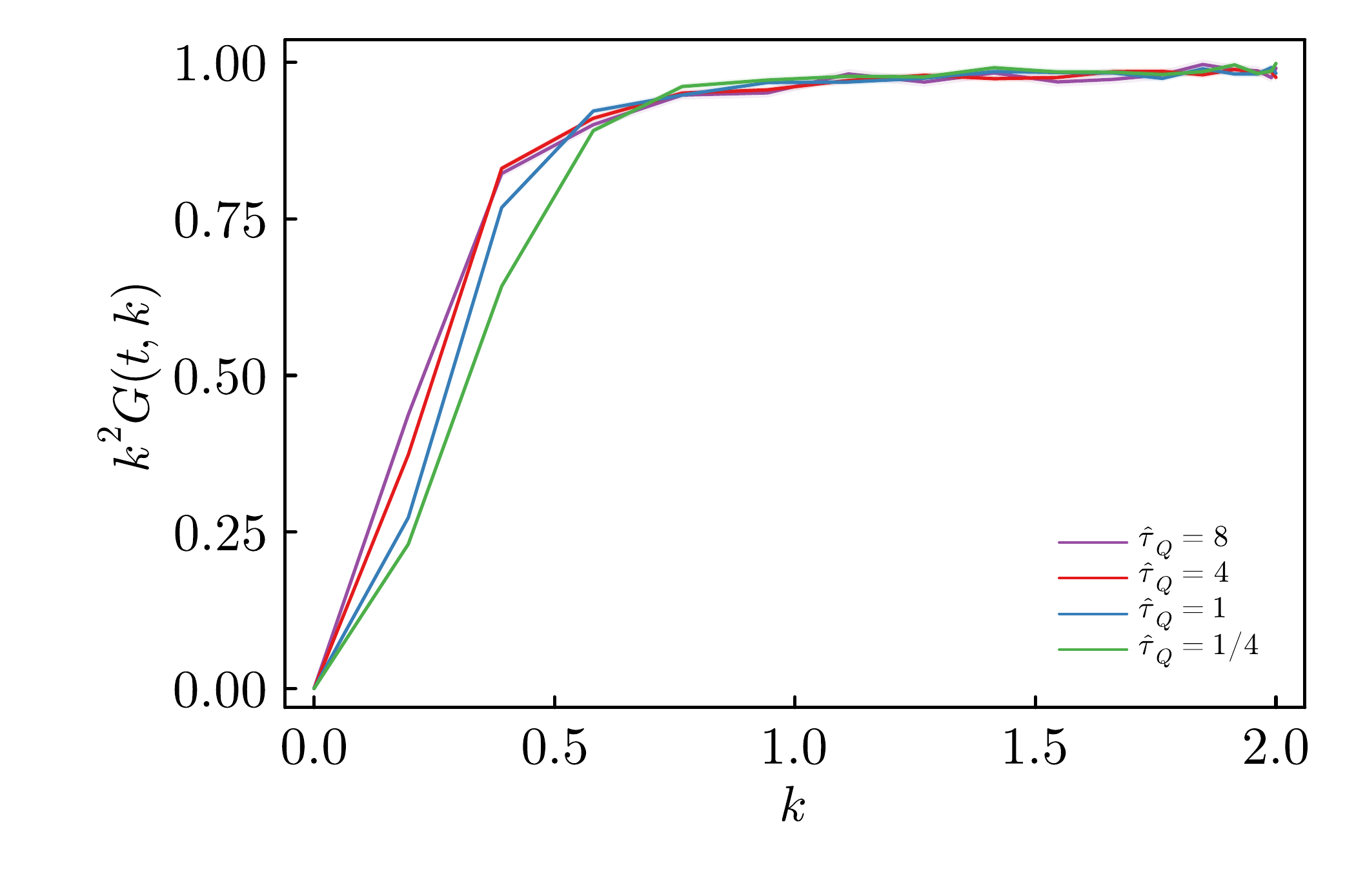}
\hspace*{0.22\linewidth}
\caption{Scaled equal time correlation functions $k^2G(t_i;0,k)$
obtained by sweeping the system from the high temperature 
phase to an instantaneous value of $m^2(t_i) = -2.2145$
(slightly above the critical value $m_c^2\simeq -2.2858$). 
We consider four different sweep times $\hat{\tau}_Q=(8,4,
1,\frac{1}{4})$, where $\hat{\tau}_Q=\tau_Q/\tau_R$. }
\label{fig:KZ}
\end{figure}

 We can define an instantaneous relaxation time $\tau(t)$ and correlation
length $\xi(t)$. Near equilibrium dynamical scaling implies $\tau(t) 
\propto  \xi(t)^{z}$. The Kibble-Zurek time can be obtained from the 
condition 
\begin{align}
    \dot \tau(\tau_{KZ}) = 1 , 
\end{align}
which expresses the condition that the rate of change of the relaxation
time is comparable to the relaxation time itself.
In the following we will consider a specific protocol for changing the 
parameters of the model. The simplest possibility is to vary $m^2$ as 
a function of $t$. This is not the most general choice; in connection
with simulating the dynamical evolution in the QCD phase diagram it 
is more appropriate to consider the evolution of both $m^2$ and $h$, 
but the details of the trajectory depend on the precise embedding of 
the critical equation of state in QCD phase diagram \cite{Parotto:2018pwx}.

\begin{figure}
\centering
\hspace*{0.10\linewidth}
\includegraphics[width=0.66\linewidth]
{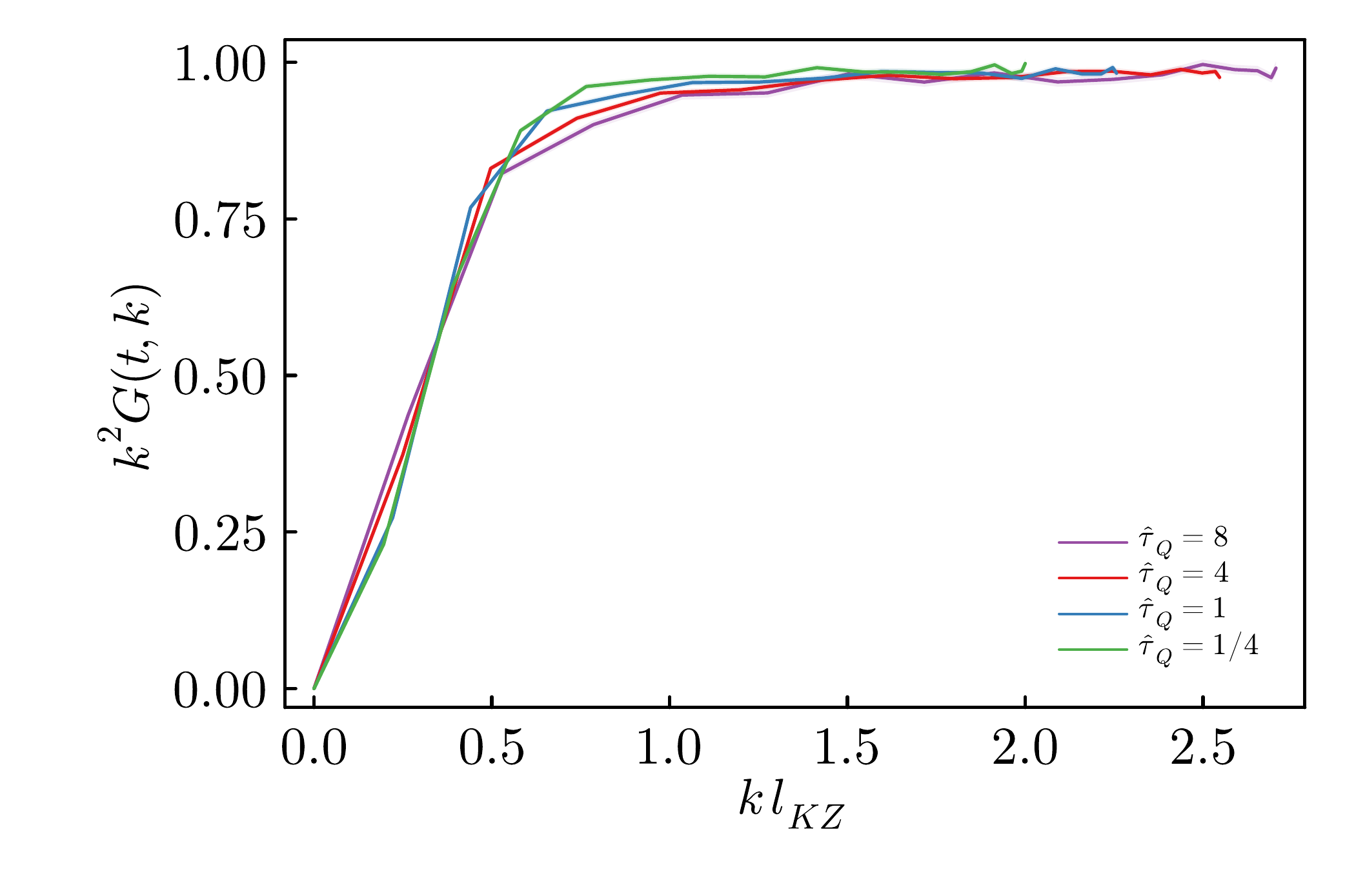}
\hspace*{0.22\linewidth}
\caption{Same as Fig.~\ref{fig:KZ} as a function of the scaling
variable $kl_{KZ}$ where $l_{KZ} \propto t_c^p$. In this figure 
we have used $p=0.08$ to achieve the best data collapse for all
values of $kl_{KZ}$. This value should be compared to the expectation 
in the scaling regime, $p=\frac{\nu}{z\nu +1}\simeq 0.16$.}
\label{fig:KZ-scaled}
\end{figure}

 Consider a power-law behavior for $m^2$ near the critical point, 
$\delta m^2 = \bar{m}^2\, | \Gamma_Q \bar{t} |^{\tilde{a}}$. Here $\delta m^2 
= m^2-m_c^2$ and $\bar{t} = t-t_c$, where $t_c$ is the critical time,
$m^2(t_c)=m_c^2$. We also defined the quench rate $\Gamma_Q$, the quench
exponent $\tilde{a}$, and an amplitude $\bar{m}^2$. We then find
\begin{align}
\dot \tau(t) \Big|_{\bar{t} = \tau_{KZ}} &
  =  \left. \frac{d}{dt} \, \xi^z(t) \right|_{\bar{t} = \tau_{KZ}}   
  =  \left. \frac{d}{dt} \, \left(\delta m\right)^{- 2z\nu}  
                          \right|_{\bar{t} = \tau_{KZ}}
  =  \left. \frac{d}{dt} \,  \left(\bar{m}\right)^{- 2z\nu } 
                  \, |\Gamma_Q \bar{t}|^{- z\nu \tilde{a}}  
                  \right|_{\bar{t} = \tau_{KZ}}
     \notag \\[0.2cm]
 & \hspace{1cm}
  = \left(\bar{m}^{2/\tilde{a}} \Gamma_Q \right)^{-z \nu \tilde{a} } 
     \left(\tau_{KZ}\right)^{-(z\nu \tilde{a}+1)}  \, , 
\end{align}
where $\nu\simeq 0.6299$ \cite{El-Showk:2014dwa} is the correlation length
exponent. This result determines the Kibble-Zurek time 
\begin{align}
\tau_{KZ} = 
    \left( \bar{m}^{2/\tilde{a}} \Gamma_Q \right)^{-
        \frac{z \nu \tilde{a}}{z \nu \tilde{a}+1}}\, . 
\end{align}
The corresponding Kibble-Zurek length is $l_{KZ} \propto \tau_{KZ}^{1/z}$.
The simplest protocol for the evolution of $m^2$ is a linear sweep 
($\tilde{a}=1$) starting at $m^2(t\!=\!0)=m_0^2$. We use 
\begin{equation}
\label{m_sweep}
    m^2(t) = m_c^2 + (m_c^2 - m_0^2) \,  \frac{t-t_c}{t_c} 
  = m_c^2 \left( 1 + \frac{t-t_c}{\tau_Q}\right) 
\end{equation}
with $\tau_Q = \Gamma_Q^{-1} = t_c m_c^2/(m_c^2 - m_0^2)$. We initialize 
the system in the symmetric phase, $m_0^2>m_c^2$. In practice we have used 
$m^2_0=-2$ (recall that $m_c^2 = -2.28587(7)$). The choice of $t_c$ then 
fixes the quench time $\tau_Q$. A scale for $\tau_Q$ is provided by the 
relaxation time $\tau_R$ of the slowest mode $k=2\pi/L$ at criticality. 
The data in Fig.~\ref{fig:cL} correspond to a relaxation time $\tau_R 
\approx 2 \cdot 10^{-3} L^z$. In order to observe Kibble-Zurek dynamics 
we need to ensure that the slowest mode is equilibrated at the beginning
of the sweep, but falls out of equilibrium as the critical point is 
approached. In the following we consider $\hat{\tau}_Q=(8,4,1,\frac{1}{4})$,
where we have defined $\hat{\tau}_Q=\tau_Q/\tau_R$.

\begin{figure}
\centering
\includegraphics[width=0.66\linewidth]{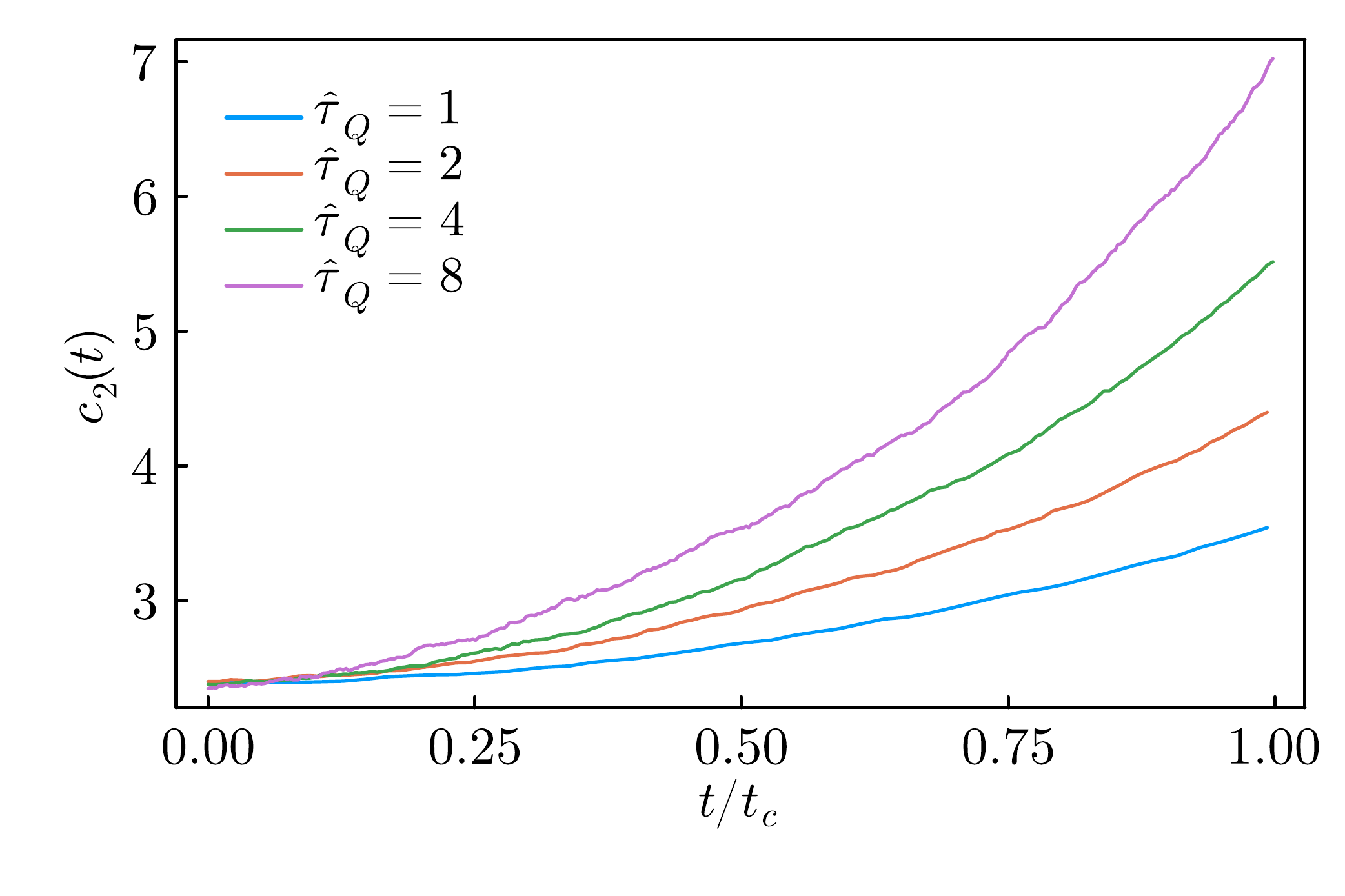}
\caption{
Time evolution of the second moment $c_2(t,\tau_Q)$ of the order parameter 
integrated over a finite volume $V=(L^3)/2$. The different curves
correspond to different sweep rates as indicated in the legend. The 
time is given in units of the critical time $t_c$ defined by $m^2(t_c)
=m_c^2$.}
\label{fig:c2_t}
\end{figure}

We observe that for a linear sweep the Kibble-Zurek time and length 
are given by
\begin{align}
\label{tau_KZ}
\tau_{KZ} &\propto \tau_Q^{\frac{z\nu}{z\nu+1}} 
                \simeq  \tau_Q^{2/3}
                \propto   t_c^{2/3}, \\
\label{l_KZ}
 l_{KZ}   &\propto \tau_Q^{\frac{\nu}{z\nu+1}} 
                \simeq \tau_Q^{1/6}
                \propto t_c^{1/6}\, . 
\end{align}
In Fig.~\ref{fig:KZ} we show the equal time correlation function of 
$\phi(t,\vec{x})$ as a function of the wave number $|\vec{k}|$ at a 
fixed instantaneous value of $m^2$ for different quench rates. A 
simple model for the equal time two-point function in equilibrium
is 
\begin{align}
 G_{\it eq}(0,k^2) = \frac{\chi_0 (\xi/\xi_0)^{\gamma/\nu}}
    {1+(k\xi)^{2-\eta}}    \, , 
\end{align}
where $\xi_0$ is the microscopic correlation length, $\chi_0$ 
the corresponding susceptibility, and $\gamma$ is the susceptibility
exponent. The scaling relation $\gamma/\nu=2-\eta$ implies that the 
asymptotic form $G(0,k\xi\gg 1)\sim k^{-2+\eta}$ is independent 
of $\xi$. We note that $\eta\simeq 0.036$ is very small, and in 
practice we have normalized the correlation function to $G_0\sim 
k^{-2}$ \footnote{Note that the value of $\eta$ is the same in model
A and B, but the large value of $z$ in model B implies that a 
numerical calculation of $\eta$ requires fewer resources in model A
than it does in model B.}. 
In a finite volume, the equal time correlation of model B is zero for 
$k=0$. This is a consequence of charge conservation, and it is not seen 
in model A simulations.

\begin{figure}
\centering
\includegraphics[width=0.66\linewidth]{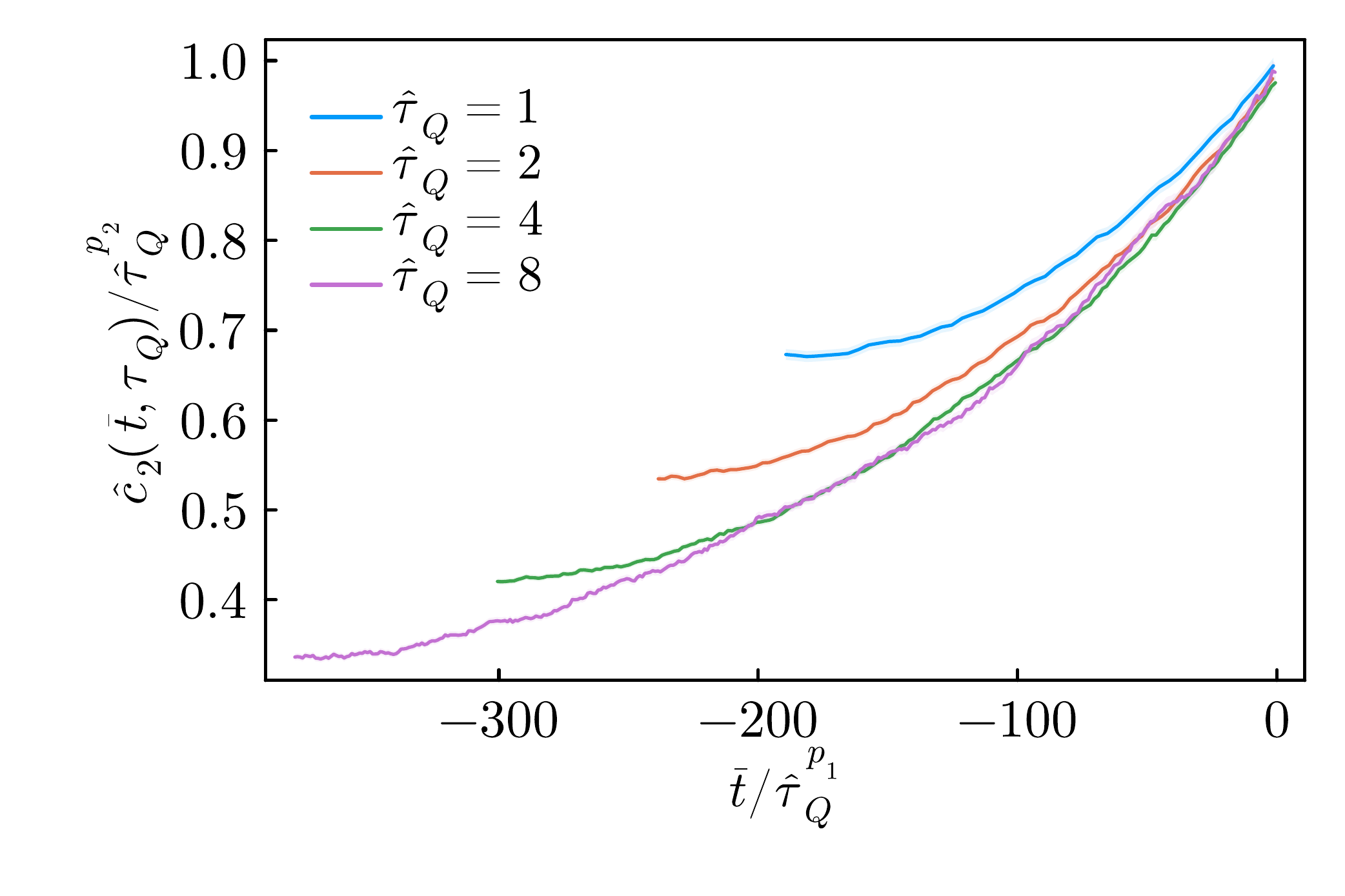}
\caption{
This figure shows the rescaled second moment $\hat{c}_2(\bar{t},\tau_Q)
/\hat{\tau}_Q^{p_2}$ as a function of $\bar{t}/\hat{\tau}_Q^{p_1}$. 
Here, we have defined $\hat{\tau}_Q=\tau_Q/\tau_R$ and the figure 
shows four different relaxation rates $\hat{\tau}_Q=1,2,4$ and 8.
In this figure we have used the mean-field critical exponents 
$p_1=2/3$ and $p_2=1/3$.}
\label{fig:c2_KZmf}
\end{figure}

 Fig.~\ref{fig:KZ} shows the correlation functions at a fixed 
instantaneous value of $m^2 =-2.2145$. This value is close to (but 
slightly larger than) the critical value $m_c^2\simeq -2.2858$.
As explained above, we consider four different values of the quench 
time, $\hat{\tau}_Q=(8,4,1,\frac{1}{4})$. We observe that the 
correlation functions do indeed fall out of equilibrium, and 
that the effect is largest for the most rapid quench. In 
Fig.~\ref{fig:KZ-scaled} we show that approximate Kibble-Zurek 
scaling holds: Data collapse is seen if $k$ is rescaled by $l_{KZ}$. 
Here we have treated the scaling exponent $p$ in the relation $l_{KZ}
\propto \tau_Q^p$ as a free parameter. The best collapse for all 
$kl_{KZ}$ corresponds to $p\simeq 0.08$, compared to the prediction
from equ.~(\ref{l_KZ}), which gives $p\simeq 0.16$. 

 We note that Kibble-Zurek scaling is based on the fact that the
correlation length is the only relevant length scale. On the lattice 
the microscopic length scale is given by the lattice spacing, and 
Kibble-Zurek scaling requires that $ka=2\pi n (a/L) \ll 1$. In 
practice, this condition is only satisfied for the lowest modes
on our lattice. Indeed, a $\chi^2$ fit to the exponent $p$, which 
emphasizes the regime of small $k$ where the statistical error is 
small, gives $p\simeq 0.16$. We conclude that a precise extraction 
of $p$ on a lattice with linear size $L=32$ is difficult, and that 
a conservative estimate of $p$ is given by $p\simeq (0.08-0.16)$.

 We have also studied the time evolution of the second moment of the 
order parameter in a finite volume during a sweep. This observable 
is related to the second cumulant that can be measured in relativistic
heavy ion collisions. We define
 \begin{align}
c_2(t,\tau_Q) = 
 \big\langle \Big( \frac{1}{V}\int_V d^3x\, \phi(x,t) \Big)^2
 \big\rangle_{\tau_Q}   
\end{align}
where we take $V=(L^3)/2$ and we average over many sweeps with 
the same quench time $\tau_Q$. The result for different sweep 
rates is shown in Fig.~\ref{fig:c2_t}. As expected, we observe
that the slowest sweep rate leads to the largest enhancement in
$c_2(t,\tau_Q)$ near $t_c$. In order to check whether these results
respect Kibble-Zurek scaling we first shift the time variable to 
$\bar{t}=t-t_c$. We also normalize $c_2(t,\tau_Q)$ to the value 
at $\bar{t}=0$ and $\tau_Q=\tau_R$, and denote the normalized
function $\hat{c}_2(\bar{t},\tau_Q)$.

We look for data collapse by plotting $\hat{c}_2(\bar{t},\tau_Q)/
\hat{\tau}_Q^{p_2}$ against $\bar{t}/\hat{\tau}_Q^{p_1}$. The 
expectation from Kibble-Zurek scaling is $p_1=\frac{z\nu}{z\nu+1}$ 
and $p_2=\frac{2\nu}{z\nu+1}$. In Fig.~\ref{fig:c2_KZmf} we show 
the evolution of the scaled moments for the mean field exponents 
$p_1=2/3$ and $p_2=1/3$. We observe approximate data collapse, but 
our data are not sufficiently accurate to distinguish between the 
mean field exponents and the prediction based on the full theory, 
$p_1=0.72$ and $p_2=0.36$.

\section{Conclusions and Outlook}
\label{sec:sum}

 We have performed numerical simulations of the diffusive dynamics of 
a conserved density (model B). The simulations are performed on a 
spatial lattice, and the time evolution is governed by a Metropolis
algorithm. This algorithm is designed such that the first moment
of the Metropolis step reproduces the diffusion equation, and the
second moment matches the variance of the stochastic force. The
Metropolis method also ensures that the equilibrium distribution
is governed by the free energy functional even if the time step
is not small. 

 The algorithm for the evolution of a conserved charge (model B)
is based on updating fluxes and satisfies global charge conservation
exactly. We have implemented this algorithm on a checkerboard which 
enables the update to be parallelized. The dynamical critical 
exponent in model B, $z\simeq 4$, is significantly larger than that 
on model A, $z\simeq 2$. This implies that it takes significantly 
longer to equilibrate or decorrelate model B configurations 
compared to model A. Based on a finite size scaling analysis 
we were nevertheless able to obtain an accurate value of the 
dynamical exponent, $z=3.972$, consistent with expectation $z=
4-\eta$ from the dynamic renormalization group \cite{Hohenberg:1977ym},
combined with a determination of $\eta$ based on the conformal
bootstrap \cite{El-Showk:2014dwa}.

  In practical applications, for example when attempting to model
the evolution of baryon number in relativistic heavy ion collisions, 
we expect the system to fall out of equilibrium as the critical 
regime is approached. We have modeled this behavior by considering 
non-equilibrium sweeps of the mass parameter in the Ising free
energy, starting from the high-temperature phase. We find evidence 
for approximate Kibble-Zurek scaling when we compare results for 
different sweep rates. This result both gives additional credence
to our numerical methods, and it supports the use of Kibble-Zurek
scale to estimate the maximum magnitude of critical fluctuations
in far-from equilibrium systems. 

  There are several possible extensions and improvement of the work 
described here. One important goal is to couple stochastic diffusion
to a fluid dynamic background that incorporates a realistic trajectory 
in the QCD phase diagram, and takes into account advection of the 
conserved density by the motion of the expanding fluid. A second
objective is to include fluctuations in the fluid velocity. This 
corresponds to considering the dynamics of model H \cite{Hohenberg:1977ym}. 
In this theory fluctuations of the order parameter couple to shear modes 
of the momentum density. Since the shear viscosity is only very weakly 
singular near the phase transition this coupling is expected to 
change the dynamical exponent to $z\simeq 3$, intermediate between
the model A result $z\simeq 2$ and the model B exponent $z\simeq 4$.

\acknowledgments
We thank Katie Newhall and Derek Teaney for useful discussions. We 
acknowledge computing resources provided on Henry2, a high-performance 
computing cluster operated by North Carolina State University. This work 
is supported by the U.S. Department of Energy, Office of Science, Office 
of Nuclear Physics through the Contracts DE-FG02-03ER41260 (T.S.) 
and DE-SC0020081 (V.S.).

\bibliography{bib}

\end{document}